\documentclass[useAMS,usenatbib]{mn2e}
\usepackage{graphicx,natbib}
\bibpunct{(}{)}{;}{a}{}{,}

\def\apj{ApJ}
\def\1p5{1.5DF}

\def\flux{\rm erg~s$^{-1}$~cm$^{-2}$}
\def\ergss{erg~s$^{-1}$}
\def\chandra{\emph{Chandra}}

\begin{document}

\sloppypar

\title[Luminosity function of faint Galactic sources in the CBF]{Luminosity function of faint Galactic sources in the Chandra bulge field}

\author[M. Revnivtsev et al.]{Revnivtsev M. $^{1,2}$ \thanks{E-mail: revnivtsev@iki.rssi.ru}, 
Sazonov S.$^{1}$, Forman W.$^{3}$, Churazov E.$^{4,1}$, Sunyaev R.$^{4,1}$\\
$^{1}$ Space Research Institute, Russian Academy of Sciences, Profsoyuznaya 84/32, 117997 Moscow, Russia \\
$^{2}$ Excellence Cluster Universe, Technische Universit\"at M\"unchen, Boltzmannstr.2, 85748 Garching, Germany\\
$^{3}$ Harvard-Smithsonian Center for Astrophysics, 60 Garden Street, Cambridge, MA 02138, USA\\
$^{4}$ Max-Planck-Institut f\"{u}r Astrophysik, Karl-Schwarzschild-str.1, 85741, Garching, Germany
}

\date{Accepted ??, Received ??}

\pagerange{\pageref{firstpage}--\pageref{lastpage}} \pubyear{2010}

\maketitle

\label{firstpage}

\begin{abstract}
We study the statistical properties of faint X-ray sources
  detected in the {\it Chandra Bulge Field}. The unprecedented
  sensitivity of the Chandra observations allows us to probe the
  population of faint Galactic X-ray sources down to luminosities 
  $L_{\rm 2-10 keV}\sim10^{30}$ \ergss~ at the Galactic
  Center distance. We show that the luminosity function of these CBF
  sources agrees well with the luminosity function of sources 
  in the Solar vicinity \citep{sazonov06}. The cumulative luminosity density 
  of sources detected in the CBF in the luminosity range $10^{30}-10^{32}$
  \ergss~ per unit stellar mass is $L_{\rm 2-10
    keV}/M_\star=(1.7\pm0.3)\times10^{27}$ \ergss~
  $M_\odot^{-1}$. Taking into account sources in the luminosity 
  range $10^{32}-10^{34}$ \ergss~ from \cite{sazonov06},
  the cumulative luminosity density in the broad luminosity range
  $10^{30}-10^{34}$ \ergss~ becomes $L_{\rm 2-10
    keV}/M_\star=(2.4\pm0.4)\times10^{27}$ \ergss~ per $M_\odot$. The
  majority of sources with the faintest luminosities should be active
  binary stars with hot coronae based on the available luminosity function
  of X-ray sources in the Solar environment.
\end{abstract}

\section{Introduction}

The hot interstellar plasma in galaxies \cite[see e.g.][]{forman85}
provides a valuable tool to probe different galactic phenomena,
including energy and mass ejection from supernova explosions
\cite[e.g.][]{loewenstein91,maiz01,strickland07}, AGN activity
\cite[e.g.][]{padovani93,forman05}, and the contribution of non-thermal
pressure in hot atmospheres\cite[][]{churazov10}.
However, before one can study the genuine X-ray emission of the hot
interstellar plasma in early type galaxies, one must excise or account
in some manner for the contribution of other constituents of galaxy
emission. Revnivtsev et al. (2008) summarized the contributions from
low mass X-ray binaries (see also \citealt{gilfanov04}) and the
fainter emission from the population of unresolved cataclysmic
variables (CVs) and coronally active binary stars (ABs) (see
Revnivtsev for a discussion of the origin of the faint diffuse emission).  

The appearance of the latest generation of X-ray observatories with fine
angular resolution (e.g., as high as $0.6''$ from the {\it Chandra} X-ray
Observatory) provides the possibility to subtract/mask the contribution
of the brightest X-ray binaries in nearby galaxies. The contribution of
fainter X-ray binaries is often estimated via extrapolation of
the observed luminosity function of bright sources towards lower
luminosities \cite[e.g.][]{sarazin01}, which Gilfanov (2004) has shown
to be a good estimator for this component. 

In the past, is has also been assumed that all emission lines at
energies $<$1-1.5 keV, visible in the spectra of galaxies, originate
in a hot interstellar plasma \cite[e.g.][]{sarazin01,irwin02,david06}. 
However, it is now clear (see
\citep{revnivtsev06,sazonov06,revnivtsev07,revnivtsev08,revnivtsev09,revnivtsev10}) that faint discrete galactic X-ray sources, namely cataclysmic
variables and coronally active (usually binary) stars,
provide an important contribution to the X-ray emission of galaxies
(after subtraction of the contribution from the brightest sources) and
this should be taken into account. Moreover, as the X-ray emission
from these stars naturally includes a contribution from optically thin
plasma of moderate or low temperatures, their spectra contain exactly
the same lines that are usually thought to originate in the
interstellar plasma.

Until now we have had very limited knowledge of the broad band
luminosity function of the  population of faint Galactic X-ray sources
(however,  there   are  studies  of   different  sub-populations,  see
e.g.  \citealt{pallavicini81,worrall83,hunsch99,schwope02}). Essentially,  the
best available broad band ($10^{27}$   \ergss$~  <L_{\rm   2-10
  keV}<10^{34}$ \ergss) luminosity function was derived from all sky
surveys of RXTE and ROSAT observatories by \cite{sazonov06}. Deep {\it
  Chandra}  observations  of   the  peculiar  Galactic  Center  region
\citep{revnivtsev07b} added strong new support to the universality
of the luminosity function over the entire Galaxy.

Now, the new 1 Ms {\it Chandra} observation of the galactic bulge
region -- the Chandra bulge field (CBF)
\cite[see][]{revnivtsev09,revnivtsev10} -- allows us to probe the
luminosity function of Galactic X-ray sources (mostly the old stellar
population because sources are located in the Galactic bulge region,
see \citealt{revnivtsev10}) to remarkably low luminosities.  In
particular, in the CBF, we can probe to $L_{\rm x}\sim10^{30}$ \ergss~
at the Galactic Center distance and thus to compare the derived
luminosity function to that derived from the Solar vicinity \citep{sazonov06}

In this paper we explore the statistical properties of the discrete
sources detected in the CBF and present their luminosity function and
cumulative emissivity.

\section{Data preparation}

\subsection{\chandra~ data}

We used all \chandra~ ACIS-I observations of the CBF (observation
ID \#5934, 6365, 6362, 9500, 9501, 9502, 9503, 9504, 9505, 9854, 9855,
9892, 9893) with total effective exposure time $\sim 898$ ks.  The
\emph{Chandra} data were reduced following the standard procedure fully
described in \cite{2005ApJ...628..655V}. The detector background was
modeled using the stowed dataset
(http://cxc.harvard.edu/contrib/maxim/stowed). In this work we
restrict our study to the circular region of $2.56$ arcmin-radius around
$l^{II}=0.113^\circ$, $b^{II}=-1.424^\circ$, close to the telescope
optical axis (the region dubbed HRES in \citealt{revnivtsev09}),
optimizing the angular resolution of the instrument and thus its
sensitivity.

To further enhance the sensitivity of the observation, we have
searched for sources in the summed image of the HRES region in the broad
0.5--7~keV energy band. The sensitivity limit is $f_{\rm lim}\sim
10^{-16}$ erg~s$^{-1}$~cm$^{-2}$. It is calculated
for the energy spectrum of photons collected within the CBF in the energy 
band 0.5--7 keV, approximated by a combination of models {\tt mekal} 
with $kT=0.6$ keV (abundance of elements was fixed at 0.1 of the Solar value) and {\tt bremss} with $kT=7$ keV from the XSPEC
package -- the simplest model which allows us to reproduce the broad band properties of the energy spectrum. Our model also includes interstellar absorption  with column density 
$N_H\sim7\times10^{21}$ cm$^{-2}$. This value of interstellar absorption 
towards the CBF is adopted based on the results of \cite{hong09},
measurements of $A_V$ in \citealt{revnivtsev09b,revnivtsev10}, and the 
$N_H/A_V$ ratio from \citealt{predehl95}.

\begin{figure}
\includegraphics[width=\columnwidth,bb=30 176 572 700,clip]{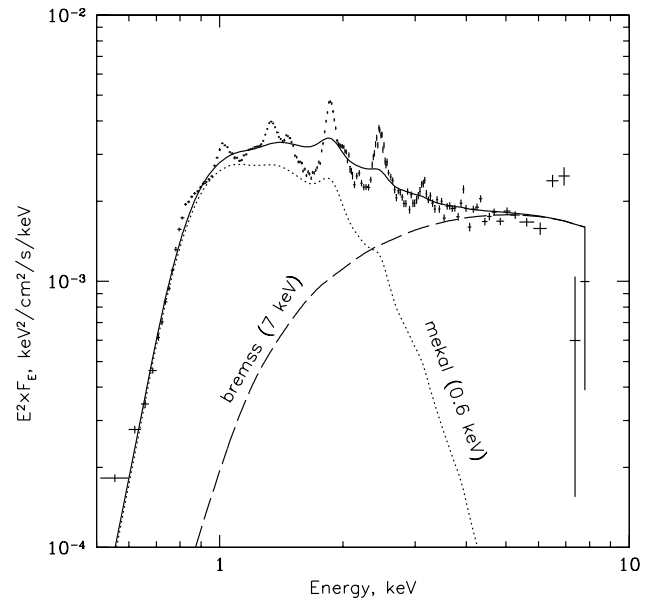}
\caption{
Energy specrum, collected by {\it Chandra} in HRES area. For conversion of {\it Chandra} counts into fluxes of sources we have adopted the simplest two component model ({\tt wabs*(mekal+bremss)} model in XSPEC package, abundance of elements in $mekal$ model is fixed at 0.1 of the Solar value), which allow us to reproduce the broad band properties of the energy spectrum. The spectral model was convolved with 100 eV Gaussian function to mimic a finite energy resolution of the instrument}
\label{total_spectrum}
\end{figure}

 Energy spectrum, collected from HRES is presented in Fig.\ref{total_spectrum}. Contribution of {\tt mekal} spectral component to total 0.5--7 keV flux is $4.83\times 10^{-12}$ \flux\ , contribution of {\tt bremss} spectral component is $3.98\times10^{-12}$ \flux\ .

The adopted sensitivity limit corresponds 
to a minimum detectable unabsorbed luminosity 
$L_{\rm 0.5-7~keV}\sim 10^{30}$ erg~s$^{-1}$ at
a source distance of 8 kpc, at which most of the Galactic objects in this
field are expected to reside. In total, we have detected 473 sources
with statistical significance $>4\sigma$ (minimum number of counts per
source is about 10). Statistical significance of detection was
calculated with the help of numerical simulations of the image taking
into account the background  and the exposure map.

Due to the significant interstellar absorption of X-rays in the CBF 
and the great diversity of spectral shapes of detected sources
(e.g., the 2--7 keV/0.5--2 keV hardness ratio of cataclysmic variables and
coronally active stars may differ by more than an order or magnitude),
it is hard to make a definitive conversion of source count rates
observed by {\it Chandra} into intrinsic fluxes and luminosities in
the broad 0.5--10 keV energy band. Therefore, despite the detection
of sources in the broad energy band 0.5--7 keV, we present here only
statistics of sources for the energy band 2--10 keV. Technically,
the source fluxes were measured by {\it Chandra} in the energy band
2--7 keV and subsequently recalculated into the 2--10 keV
energy band assuming the simple analytic form of the spectral fit
described above, appropriate for the summed spectrum of all
sources. We find that 1 count in the range 2--7 keV per image
(898 ks effective live time) corresponds to a source flux
$4.6\times10^{-17}$ \flux\ in the energy band 2--10 keV.

\subsection{Stellar mass profile along the line of sight}

According to statistics of X-ray sources in the Solar vicinity
\citep{sazonov06}, the majority of sources with luminosities $L_{\rm
  2-10 keV}>10^{30}$ \ergss~ are old binaries, including magnetic and non-magnetic cataclysmic variables, RS CVn systems, Algol-type binaries, W UMa-typa binaries. This allows
us to safely assume that their density distribution is directly
proportional to the stellar mass density distribution in the
CBF. Therefore, it is useful to obtain the luminosity function of
sources in the CBF normalized to the total mass of stars within 
the studied volume.

The stellar density distribution along the line of sight in the CBF
was previously studied in \cite{revnivtsev10}. The main substructures
of the Galaxy visible in the CBF are the Galactic disk, the Galactic
bulge and the nuclear stellar disk (NSD). The majority of sources
reside in the Galactic bulge and are located at $\sim$8 kpc. In
\cite{revnivtsev10}, it was shown that the rms-width of the apparent
magnitude distribution of red clump giants (RCG; considered to be
almost standard candles) is approximately 0.28$^m$ (this rms scatter
includes the intrinsic scatter of the RCG population, therefore the
scatter caused by the distance variations alone is even smaller than
0.28$^m$). This means that the error in source luminosities, 
associated with the assumption that all sources in the CBF are located
at 8 kpc, is less than $\sim30\%$.

To determine the influence of these distance variations on the
resulting luminosity function, and on our estimates of the luminosity
density of sources, we have simulated a set of sources assuming that
their intrinsic luminosity function has the shape $dN/dL\propto
L^{-2}$ (which is close to what is actually observed) and distributed
the sources along the line of sight according to the stellar density
distribution, described in \cite{revnivtsev10}. The resulting
luminosity function does not change by more than 10\% and the
luminosity density changes by only $\sim5\%$ in comparison
to the case of constant source distances.

According to the simple model of the stellar mass distribution 
(including the bulge, disk and the nuclear disk components) adopted in 
\cite{revnivtsev10}, the total stellar mass
angular surface density in CBF is $\sim4.6\times10^{4}~M_\odot$
arcmin$^{-2}$. We corrected for the fact that this stellar surface
density corresponds to a $\sim13\%$ higher NIR surface brightness than
that measured by {\it Spitzer} in this area \citep{revnivtsev10}. 
Therefore in our subsequent analysis we adopt
a stellar surface density $4.1\times10^{4}~M_\odot$ arcmin$^{-2}$ in
the CBF and assume that the vast majority of RCGs lie at a distance 8
kpc.  This surface mass density is obtained via integration of the adopted density profile up to further edge of the Galaxy, i.e. up to $\sim$20 kpc distance from the Sun. Approximately 85\% of the all mass contained in small solid angles in the direction of CBF is located within 1 kpc of the Galactic Center.

\section{Luminosity function}

\begin{figure}
\includegraphics[width=\columnwidth]{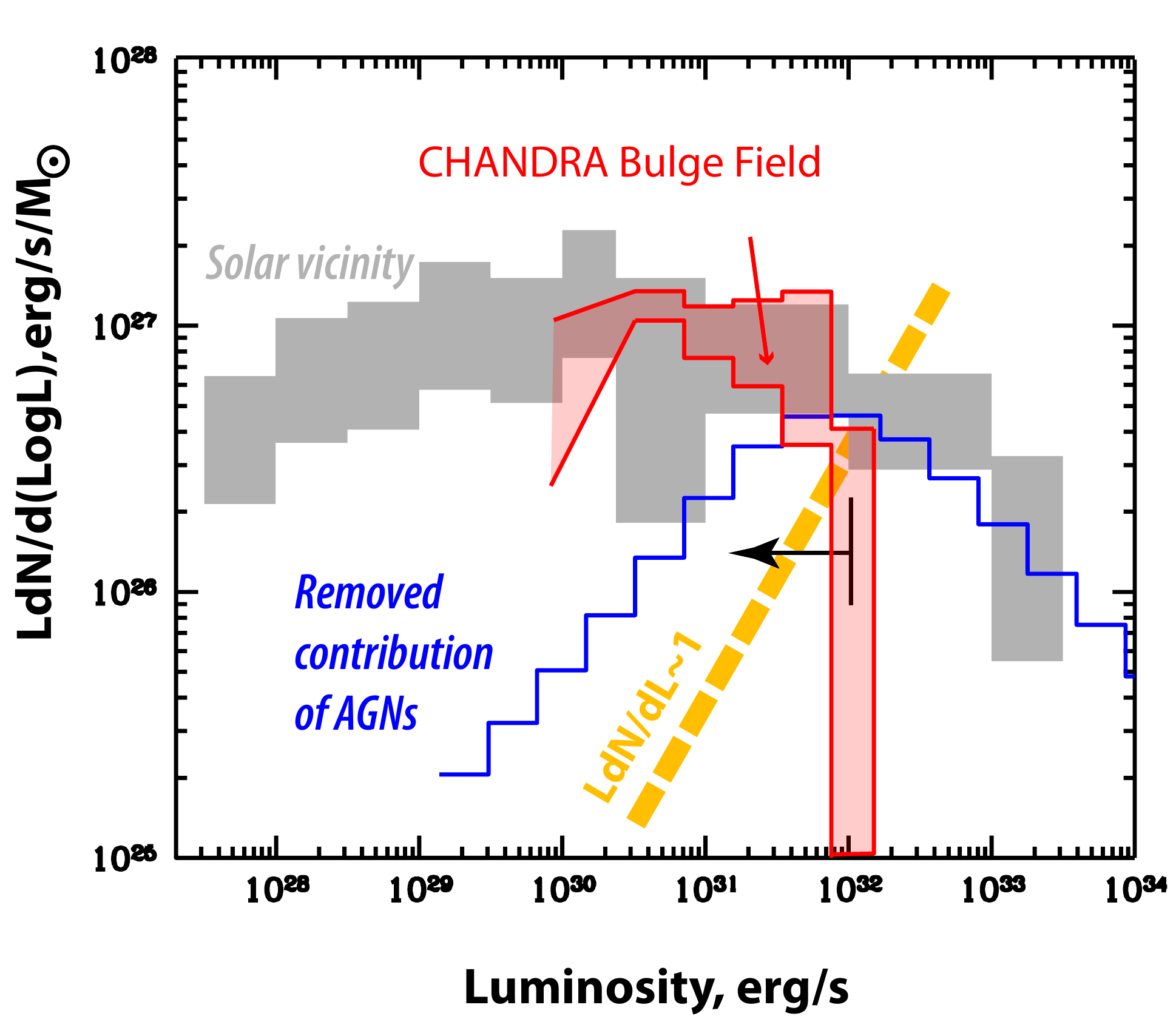}
\caption{
  The 2--10 keV luminosity function of sources detected in the \chandra~
  Bulge Field, corrected for the contribution of background active galactic 
  nuclei. Numbers of sources, contributing to the presented bins of the 
  luminosity function are 1,7,15,31,76,103 from the highest to the lowest presented bin. For comparison we also present the luminosity function of faint 
  sources obtained from the RXTE and ROSAT all sky surveys from 
  Sazonov et al. (2006). The blue histogram shows the estimated (and
  subtracted) contribution of extragalactic sources in the CBF. The area
  unavailable to our survey due to the small number of sources per
  field of view is denoted by a dashed line $dN/d\log L \sim 1$. The
  luminosity of the brightest sources in our survey is marked by an arrow.}
\label{lumfunc}
\end{figure}

In calculating the number-flux function (or luminosity function) of
sources, we confined our analysis to the flux range where we expect no
strong influence from incompleteness effects, i.e., where we detect
more than $\sim10$ counts from sources in the energy band used
for source detection (0.5--7 keV) and more than 5 counts from sources
in the energy band 2--10 keV. In principle, it is possible to continue
the luminosity function toward lower luminosities, but accounting for
the variety of spectral shapes at these luminosities (cf. soft spectra of
coronally active stars and hard spectra of dwarf novae) introduces significant
uncertainties -- the same number of counts in total energy band, used for detection of sources, (0.5-7 keV) might correspond to significantly different energy flux in 2-10 keV energy band.

The number-flux function in its differential form, normalized to the
mass of stars in the CBF area ($4.1\times10^{4}~M_\odot$ arcmin$^{-2}$), 
is presented in Fig.~\ref{lumfunc}.

To account for the contribution of extragalactic sources (mostly AGNs)
to the number-flux function detected in the CBF, we have adopted the
results of \cite{moretti03}. We calculated the anticipated number of
AGNs in every luminosity bin, assuming the Moretti et al. analytic
form of the AGN number flux function (in the energy band 2--10
keV). The AGN contribution has its own uncertainty mainly caused by
either Poisson noise of the number of sources in the field, or by cosmic
variance. According to simple estimates of these uncertainties from
the COSMOS survey \citep{cappelluti07}, in the small HRES field,
the contribution from Poisson noise of the number of
sources should dominate. We see that this noise is comparable with the
Poisson noise of the number of Galactic sources in the field.

The shape of the luminosity function is not clear, however it can be approximated more or less adequately by a simple power law in luminosity interval $30.5<\log L_x<32$: $LdN/d \log L\approx A ({L_x/{10^{31} erg/sec}})^{-\alpha}$, where $A\approx1.1\times10^{27}$ erg/s/$M_\odot$, and $\alpha=0.2\pm0.1$

\begin{table}
\caption{Cumulative luminosity densities of Galactic sources as a
  function of luminosity in the energy band 2--10 keV. Note that due to the
  limited survey volume we have only sources with luminosities
  less than $10^{32}$ \ergss.}
\begin{tabular}{c|c}
  Lum. range &Cumulative lum. density,\\
  ($>\log L_{\rm x}$, $\log L_{\rm x}<32$) & $10^{27}$erg/s/$M_\odot$\\
  \hline
  31.5& $0.49\pm0.25$\\
  31.1& $0.81\pm0.27$\\
  30.7& $1.2\pm0.3$\\
  30.3& $1.6\pm0.3$\\
  29.9& $1.7\pm0.3$\\
  \hline
  29.9--34.0$^{*}$&$2.4\pm0.4$\\
\end{tabular}
\begin{list}{}{}
\item $^*$ -- This cumulative luminosity density includes contribution
  from sources with $L_{\rm 2-10 keV}>10^{32}$ \ergss~, adopted
  from \cite{sazonov06}
\end{list}
\label{table_cumlum}
\end{table}

\begin{figure}
  \includegraphics[width=\columnwidth]{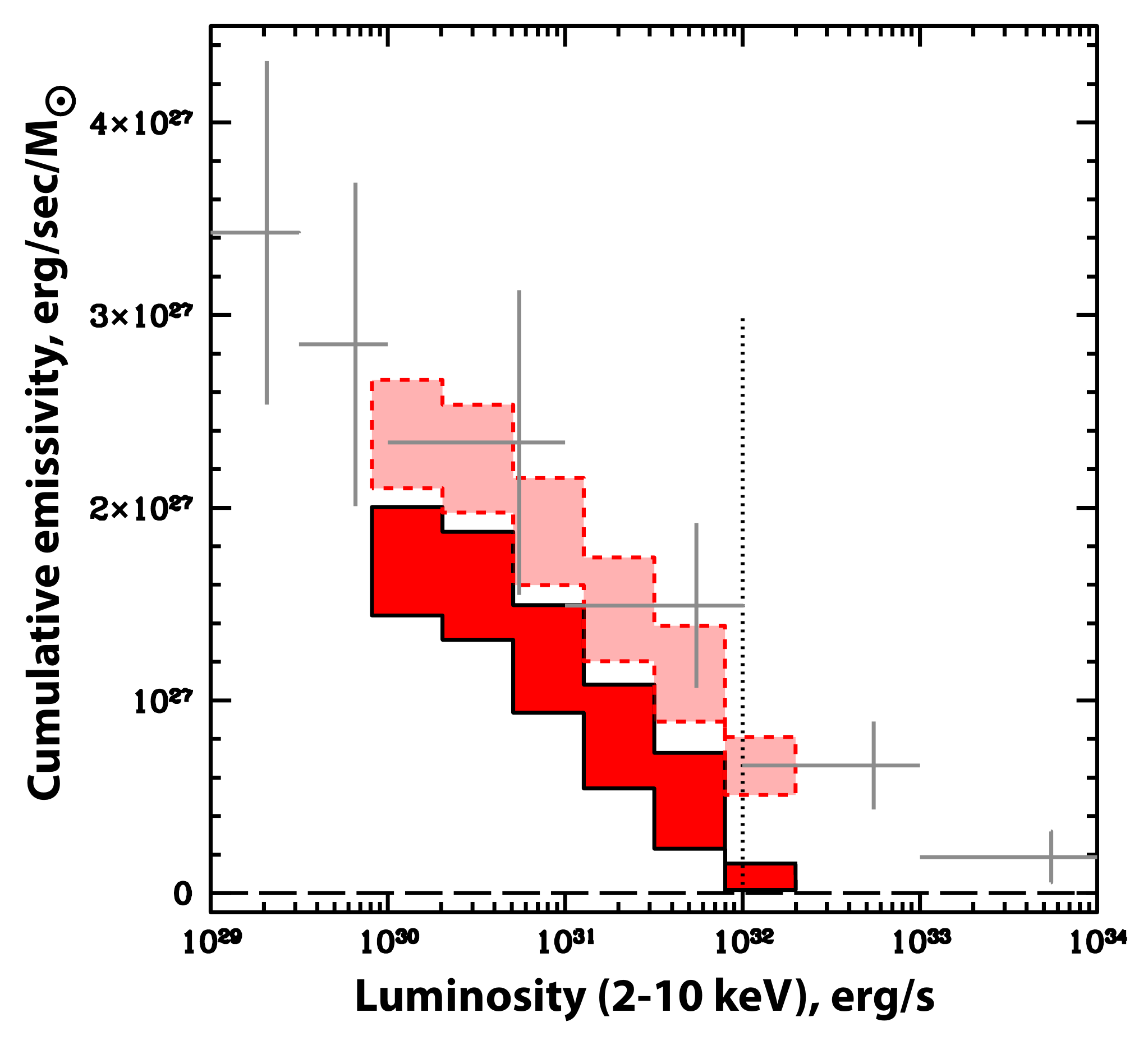}
  \caption{
  Cumulative unit stellar mass emissivity in the 2--10 keV energy band 
  of the population of Galactic X-ray sources detected in the CBF (red
  area). The estimated contribution of AGNs was subtracted (see text
  for details). Since we do not probe sources with $L_{\rm 2-10%
   keV}>10^{32}$ erg s$^{-1}$ (indicated by the vertical dotted
  line), we have added the cumulative emissivity of sources at 
  $L_{\rm 2-10 keV}>10^{32}$ \ergss~ from Sazonov et al. (2006). The resulting
  shaded area with dashed perimeter can be compared with the corresponding 
  result from Sazonov et al. 2006 (grey crosses).}
\label{cumlum}
\end{figure}

The cumulative emissivity of Galactic sources as a function of the
limiting X-ray luminosity is shown in Table~\ref{table_cumlum} and
Fig.~\ref{cumlum} (after correcting for AGNs as described above). Due 
to the relatively small survey volume, we cannot probe
Galactic sources with luminosities higher than $10^{32}$ \ergss~
(increasing the solid angle of our study will not help here because
the number of extragalactic sources will become a significant
contributor to the source counts). Therefore, our estimates of the
cumulative emissivity of sources lack the contribution from higher
luminosity sources (in particular, luminous cataclysmic variables).
To compare our results with the luminosity density of sources
determined in the Solar vicinity \citep{sazonov06}, we added the value
of the cumulative emissivity measured for sources with $10^{32}$ \ergss~
$<L_{\rm 2-10 keV}<10^{34}$ \ergss~ (shaded area with dashed perimeter
in Fig.~\ref{cumlum}). The resulting cumulative luminosity density of
sources with $10^{30}$ \ergss~ $<L_{\rm 2-10 keV}<10^{34}$ \ergss~ is
$L_{\rm 2-10 keV}/M_\star=(2.4\pm0.4)\times10^{27}$ \ergss per $M_\odot$.

\subsection{Contributions from various stellar populations}

It is practically impossible to determine the nature of all Chandra
sources due to the extremely high spatial density of stars in the
direction of the CBF \cite[see e.g. images
  in][]{revnivtsev10}. However, we can estimate the composition
judging from our knowledge of the luminosity function of different
types of sources in the Solar vicinity. According to the space densities 
of different types of sources presented in \cite{sazonov06}, we can anticipate 
approximately equal contributions from cataclysmic variables and from 
coronally active binaries to the cumulative 2--10~keV emisison in the 
luminosity range $10^{31}$ \ergss $<L_{\rm 2-10 keV}<10^{32}$ \ergss, 
and a dominant contribution of ABs at energies below 2~keV.

\begin{figure}
\includegraphics[width=\columnwidth,bb=30 185 567 700,clip]{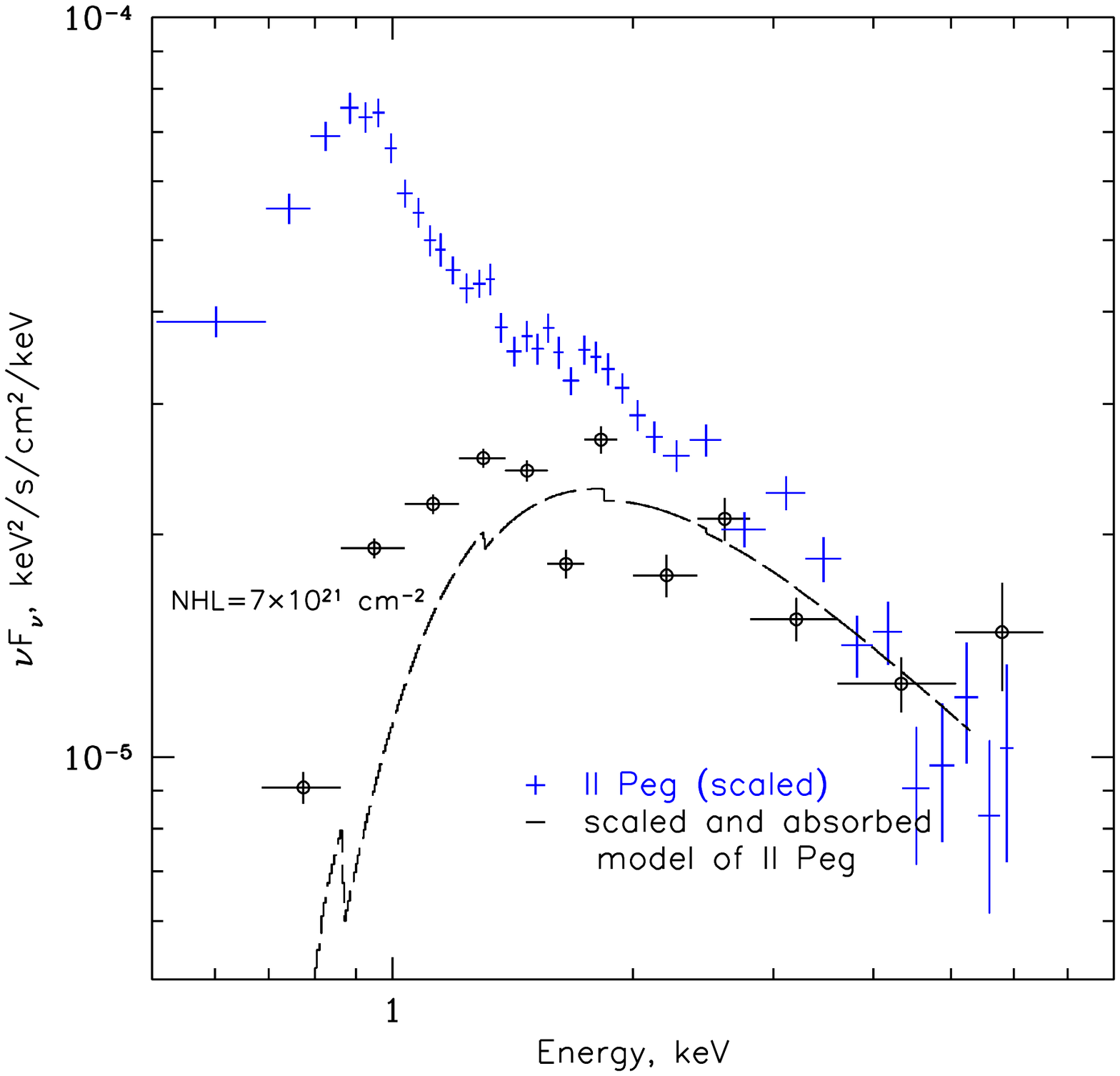}
\caption{Spectrum of the faintest sources (with 5--10 net counts per source 
  over the whole exposure time, i.e. $L_{\rm x}\sim 10^{30}$ \ergss),
  detected in the CBF (open circles).  For comparison we present 
  the spectrum of a typical bright active binary star II Peg (ASCA
  Observatory data, observations Dec. 18--19, 1994), scaled
  to match the {\it Chandra} spectrum at energies 3--7 keV. The dashed
  curve shows a simple spectral model of a broken power law, which was
  fit to the spectrum and then absorbed with
  $N_H=7\times10^{21}$ cm$^{-2}$, appropriate to the CBF. The curve has arbitrary
  normalization. It can be seen that the spectral shape of the faintest {\it
    Chandra} sources is compatible with that of bright ABs.}
\label{spectra}
\end{figure}

Unfortunately, measurements of the densities of cataclysmic variables at lower
luminosities ($L_{\rm 2-10 keV}<10^{31}$ \ergss) are more
uncertain. Judging from estimates presented in \cite{pretorius07} for 
CVs in the North Ecliptic Pole region (Table 2 of
this paper), their spatial density in the range $10^{30}\textrm{
  \ergss}<L_{\rm 2-10 keV}<10^{31}$ \ergss~ should be $\sim1.6\times
10^{-6}$ pc$^{-3}$, or $\sim4\times10^{-5}$ $M^{-1}_\odot$, assuming a
local stellar density of $\sim0.04~M_\odot~pc^{-3}$
\citep{robin03}. The luminosity density of CVs can then be estimated to be 
$\sim1.2\times10^{26}$ \ergss~ per $M_\odot$. This is just $\sim 15$\%
of the luminosity density of sources with $10^{30}\textrm{
  \ergss}<L_{\rm 2-10 keV}<10^{31}$ \ergss~ in the CBF ($\sim
  (8-9)\times10^{26}$ \ergss~ per $M_\odot$). This
suggests that most of the detected sources in this luminosity range
are active binary stars.

Active binary stars produce softer X-ray emission than do cataclysmic
variables in general, so that $L_{\rm 0.5-2 keV}/L_{\rm 2-10 keV}>1$ for them (see e.g. \citealt{heinke05,sazonov06}, see \citealt{guedel04} for review of X-ray emission of AB sources). This is of high importance for soft X-ray studies of other galaxies. In
Fig.~\ref{spectra}, we show the cumulative spectrum of the faintest
sources detected in the CBF (which is clearly softer than the summed spectrum of all detected sources) and compare it to the energy spectrum of
the typical bright active binary star II Peg (see more details on spectra of II Peg in \citealt{mewe97}). The dashed curve in the
figure shows a simple broken power law fit to the spectrum of II Peg,
arbitrarily scaled and absorbed with interstellar column density
$N_H=7\times10^{21}$ cm$^{-2}$, appropriate for the CBF. We see that
the spectrum of the faintest CBF sources is compatible with that of AB
stars of the appropriate luminosity. However, the considerable
absorption at energies $<2$~keV prevents seeing the full luminosity
density of such soft sources. Without interstellar absorption, the CBF
brightness at energies $0.5-2$ keV would be several times higher. 

\section{Summary}

We have analysed statistical properties of X-ray sources detected in
deep observations of the {\it Chandra} bulge field. The results of
this work can be summarized as follows: 

\begin{itemize}
\item The luminosity function of detected X-ray sources in the luminosity
  range probed by our observations ($30<\log L_{\rm 2-10 keV}<32$)
  and corrected for the contribution of extragalactic sources (AGNs), is
  relatively flat ($L^2dN/dL\sim$ constant) and consistent with that
  determined from a sample of sources in the Solar vicinity by
  \cite{sazonov06}. 
\item The cumulative luminosity density of sources in the luminosity
  range $30<\log L_{\rm 2-10 keV}<32$ is $(1.7\pm0.3)\times10^{27}$
  \ergss per $M_\odot$. Accounting for the contribution from sources with 
  $\log L_{\rm 2-10 keV}>32$ from \cite{sazonov06}, the cumulative
  luminosity density of sources with $\log L_{\rm 2-10 keV}>30$ is
  $(2.4\pm0.4)\times10^{27}$ \ergss~ per $M_\odot$. 
\item The majority of the faintest sources should be active binaries
  with hot coronae, which makes their cumulative emission virtually
  indistinguishable from that of the hot interstellar plasma typically
  associated with old stellar populations. This
  indicates the importance of carefully accounting for the emission of such
  sources, based on detailed knowledge of their statistical properties
  and not only on the shape of the energy spectrum.
\end{itemize}

\section*{Acknowledgements}
This research made use of data obtained from the High Energy
Astrophysics Science Archive Research Center Online Service, provided
by the NASA/Goddard Space Flight Center. This work was supported by a
grant of Russian Foundation of Basic Research 10-02-00492-a, grant
NSh-5069.2010.2, and programs of the Russian Academy of Sciences P-19
and OFN-16. SS acknowledges the support of the Dynasty Foundation.

\label{lastpage}

\end{document}